# Computational study on the half-metallicity in transition metal–oxide-incorporated 2D g-$C_3N_4$ nanosheets

Qian Gao (高乾)[1], Huili Wang (王会丽)[1], Lifu Zhang (张丽芙)[1], Shuanglin Hu (胡双林)[2,*],
Zhenpeng Hu(胡振芃)[1,†]

[1]School of Physics, Nankai University, Tianjin 300071, China
[2]Institute of Nuclear Physics and Chemistry, China Academy of Engineering Physics, Mianyang, Sichuan 621900, China

Corresponding author. Email: [†]zphu@nankai.edu.cn; [*]hushuanglin@caep.cn

In this study, based on the first-principles calculations, we systematically investigated the electronic and magnetic properties of the transition metal–oxide-incorporated 2D g-$C_3N_4$ nanosheet (labeled $C_3N_4$–TM–O, TM = Sc–Mn). The results suggest that the TM–O binds to g-$C_3N_4$ nanosheets strongly for all systems. We found that the 2D $C_3N_4$–TM–O framework is ferromagnetic for TM = Sc, Ti, V, Cr, while it is antiferromagnetic for TM = Mn. All the ferromagnetic systems exhibit the half-metallic property. Furthermore, Monte Carlo simulations based on the Heisenberg model suggest that the Curie temperatures ($T_c$) of the $C_3N_4$–TM–O (TM = Sc, Ti, V, Cr) framework are 169 K, 68 K, 203 K, and 190 K, respectively. Based on Bader charge analysis, we found that the origin of the half-metallicity at Fermi energy can be partially attributed to the transfer of electrons from TM atoms to the g-$C_3N_4$ nanosheet. In addition, we found that not only electrons but also holes can induce half-metallicity for 2D g-$C_3N_4$ nanosheets, which may help to understand the origin of half-metallicity for graphitic carbon nitride.

## 1 Introduction

Tremendous theoretical, experimental, and technological efforts have been made in the last few years to understand the characteristics of half-metallic materials, particularly in regard to the properties of pure spin generation and transport[1-6]. Recently, some research has been devoted to the potential half-metallic property of g-$C_3N_4$, which is a polymeric semiconductor with a bandgap of 2.7 eV and has attracted the wide interest of researches for photocatalysis[7-9] applications.

Owing to their low-dimensionality and electron confinement, nanosheet materials have been explored as potential materials for high-performance spintronic devices[10-11]. The g-$C_3N_4$ nanosheet has been widely synthesized experimentally[12-15], and several ways have been proposed to introduce magnetism and half-metallicity for it, including doping[16-20], defecting[21], and forming nanocomposites with other materials[22]. However, there are still issues pertaining to these proposed methods in practical applications. For example, the planar $C_4N_3$ material, which can be regarded as the C-doped g-$C_3N_4$, was predicted to be half-metallic based on first-principles calculations[23]. However, the experimental synthesis of pure $C_4N_3$ is still a challenge[24]. Although introducing transition metal (TM) atoms as a source of spin moment[25-29] has also been proposed in low-dimensional systems to obtain the half-metallicity, TM-incorporated nanosheets have a major limitation in that the TM atoms on top of the layer are quite mobile in nature and easily form clusters due to strong $d$–$d$ interactions, somewhat preventing practical applications in spintronics[30-31]. Fortunately, Wu et al. successfully synthesized molecular titanium–oxide-incorporated 2D g-$C_3N_4$[32] recently, and this suggests that TM–oxide (TM–O) would be more stable on g-$C_3N_4$ nanosheets. It is necessary to study the electronic properties of TM–O-incorporated 2D g-$C_3N_4$ to determine the presence of half-metallicity.

In this study, inspired by the molecular titanium–oxide-decorated 2D g-$C_3N_4$, we investigated the half-metallicity of $C_3N_4$–TM–O (TM = Sc, Ti, V, Cr, Mn) systems. With the aim of application of this material to spintronics, we considered the important magnetic properties for half-metallic samples, such as spin polarization at Fermi energy and Curie temperature ($T_c$). Furthermore, based on the charge transfer analysis, we also explain the half-metallic origin of molecular TM–O-decorated g-$C_3N_4$ monolayers, and this provides a potential way to engineer the electronic and magnetic properties of 2D nanomaterials.

## 2 Computational method

First-principles calculations were performed based on spin-polarized density functional theory (DFT) using the generalized gradient approximation (GGA) in the form proposed by Perdew, Burke, and Ernzerhof (PBE)[33], as implemented in the Vienna *ab initio* simulation package (VASP) code[34-35]. The projector augmented-wave (PAW) method with a plane-wave basis set was used[36]. To better describe strongly correlated systems containing partially filled $d$ subshells, the GGA+U method was used in this work[37]. A correlation energy (U) of 4 eV and an exchange energy (J) of 1 eV were used for all the TM $d$ orbitals, which have been used and tested in previous theoretical works[19-20, 22, 27, 38]. We applied periodic boundary conditions and a vacuum space exceeding 16 Å along the $z$ direction to avoid interactions between two g-$C_3N_4$ nanosheet images in neighboring unit cells. All the structures were relaxed using the conjugated gradient method without any symmetric constraints. As shown in Fig. 1, there were four pores in the unit cell of g-$C_3N_4$, and only one group of TM-O was inserted into a unit cell. To investigate the magnetic coupling between the TM atoms, a 2 × 2 supercell including 16 pores was used. Monkhorst–Pack k-point meshes of 5 × 5 × 1 and a single Γ-point were used to represent their reciprocal spaces for the unit cell and the 2 × 2 supercell of $C_3N_4$–TM–O systems, respectively. The energy cutoff, as well as the convergence criteria for energy and force, was set to be 520 eV, $1 \times 10^{-5}$ eV, and 0.02 eV/Å, respectively. To determine the structural stability of these structures, the binding energies were calculated with the following formula,

$$E_b = E_{sheet-TM-O} - (E_{sheet} + E_{TM} + \frac{1}{2}E_{O_2}) , \qquad (1)$$

where $E_{sheet}$, $E_{O_2}$, and $E_{TM}$ are the energies of the pure g-$C_3N_4$ nanosheet, the oxygen molecule, and the isolated TM atom, respectively. $E_{sheet-TM-O}$ denotes the total energy of the g-$C_3N_4$ nanosheet with molecular TM–O embedded within it. To study the magnetic ordering at finite temperature, Monte Carlo simulations based on the Heisenberg model were performed. The Heisenberg Hamiltonian can be expressed as follows,

$$H = \sum_{ij} J_{ij} \times S_i S_j , \qquad (2)$$

where $i$, $j$ are the nearest-neighbor magnetic sites and $J_{ij}$ is the exchange coupling constant between them[39-40]. $S_{i(j)}$ is the total spin ($S_{tot}$) per unit at the $i(j)th$ magnetic center of the 2D nanosheet. The calculation used a supercell of 48 × 48 sites, and the magnetic moments on each site changed randomly according to the spin states.

## 3 Results and discussion

### 3.1 Structure of $C_3N_4$–TM–O

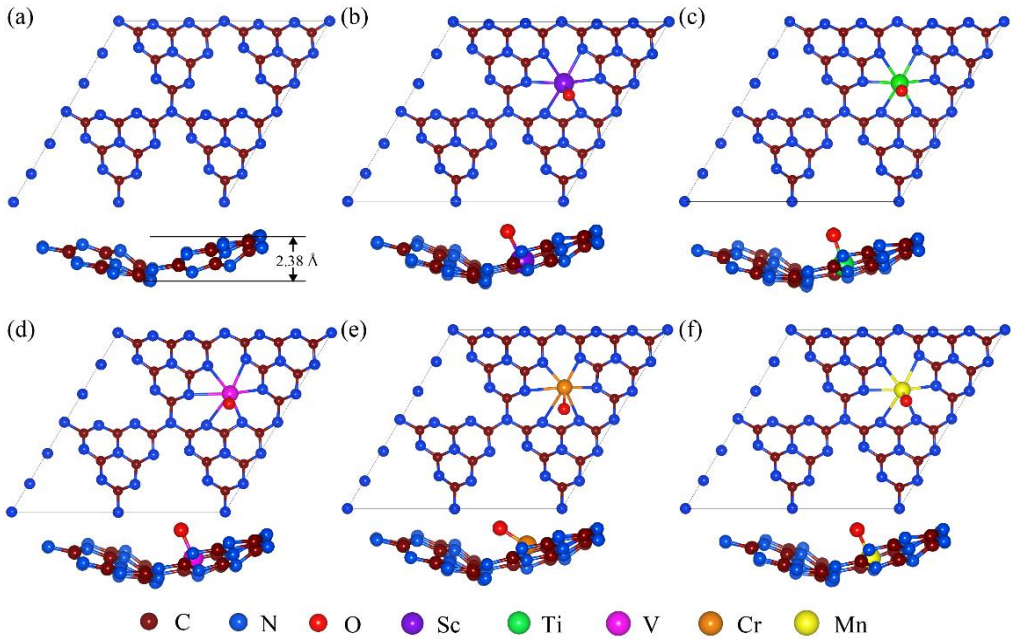

**Fig. 1** Top and side views of (a) 2D g-$C_3N_4$ structure and $C_3N_4$-TM-O systems, where TM= (b) Sc, (c) Ti, (d) V, (e) Cr, (f) Mn.

The metal-free buckled monolayer g-$C_3N_4$ is explored briefly to understand the general geometric and electronic properties. Interestingly, although the coplanar structure of the g-$C_3N_4$ nanosheet with a lattice constant of 7.14 Å has been widely used[18-19, 22], structural relaxation indicates that the nanosheet could be easily distorted due to the repulsion between lone-pair electrons of nitrogen atoms[41]. Through a linear scan of the lattice constant using a 2×2 supercell, we obtained the corrugated g-$C_3N_4$ nanosheet with a lattice constant of 6.83 Å, which is much more consistent with the experimental value of 6.81 Å[7]. The corrugated g-$C_3N_4$ is lower in energy by approximately 0.80 eV per unit cell than the coplanar structure. As shown in Fig. 1(a), the vertical distance between the highest and lowest atoms of the corrugated structure is 2.38 Å, which is quite similar to the experimental value of averaged 2.6 Å measured by atomic force microscopy[42]. The PBE-calculated energy band indicates that the corrugated g-$C_3N_4$ nanosheet is a semiconductor with a direct bandgap of 1.93 eV. All the results suggest that the corrugated g-$C_3N_4$ nanosheet with a lattice constant of 6.83 Å is more reasonable as the basis of our research.

When TM–O molecules are introduced to the cavity of g-$C_3N_4$ with four pores, 2D porous structures with regularly binding $3d$ TM–O molecules are formed. The optimized structures with the top and side view of the g-$C_3N_4$ and $C_3N_4$–TM–O systems are shown in Fig. 1. The structures of these $C_3N_4$–TM–O systems are consistent with the original structure, which also confirms the structural stability of the metal-free g-$C_3N_4$ nanosheet. The bond lengths of TM–O are 1.77, 1.68, 1.65, 1.67, and 1.69 Å for Sc, Ti, V, Cr, and Mn, respectively. As shown in Fig. 1, the TM atoms bond with the edged N atoms of the pore in g-$C_3N_4$. The distance between TM and N atoms ranges from 2.17 to 3.24 Å. These data, along with the calculated binding energies of TM, O, and $C_3N_4$ are listed in Table 1. It can be seen from Table 1 that for all the $C_3N_4$–TM–O systems, the binding energies are quite negative. For the $C_3N_4$–Sc–O and $C_3N_4$–Ti–O systems, the binding energies are even lower than -7.0 eV. These negative binding energies indicate that the TM–O molecules bind to the $C_3N_4$ system strongly. The experimental synthesis of $C_3N_4$–Ti–O further confirmed the reasonability of the structural design[32].

**Table 1**. Binding energy, magnetic moment, bond length of TM-O ($d_{TM-O}$), and distance between TM and N atoms ($d_{TM-N}$).

| Metal oxide | Sc-O | Ti-O | V-O | Cr-O | Mn-O |
|---|---|---|---|---|---|
| Binding Energy ($eV$) | -7.77 | -7.33 | -5.73 | -3.67 | -3.35 |
| Magnetic moment ($\mu B$) | 1.0 | 2.0 | 3.0 | 4.0 | 5.0 |
| $d_{TM-O}$ (Å) | 1.77 | 1.68 | 1.65 | 1.67 | 1.69 |
| $d_{TM-N}$ (Å) | 2.41-2.56 | 2.32-2.45 | 2.21-2.89 | 2.17-3.24 | 2.25-2.83 |

## 3.2 Electronic properties

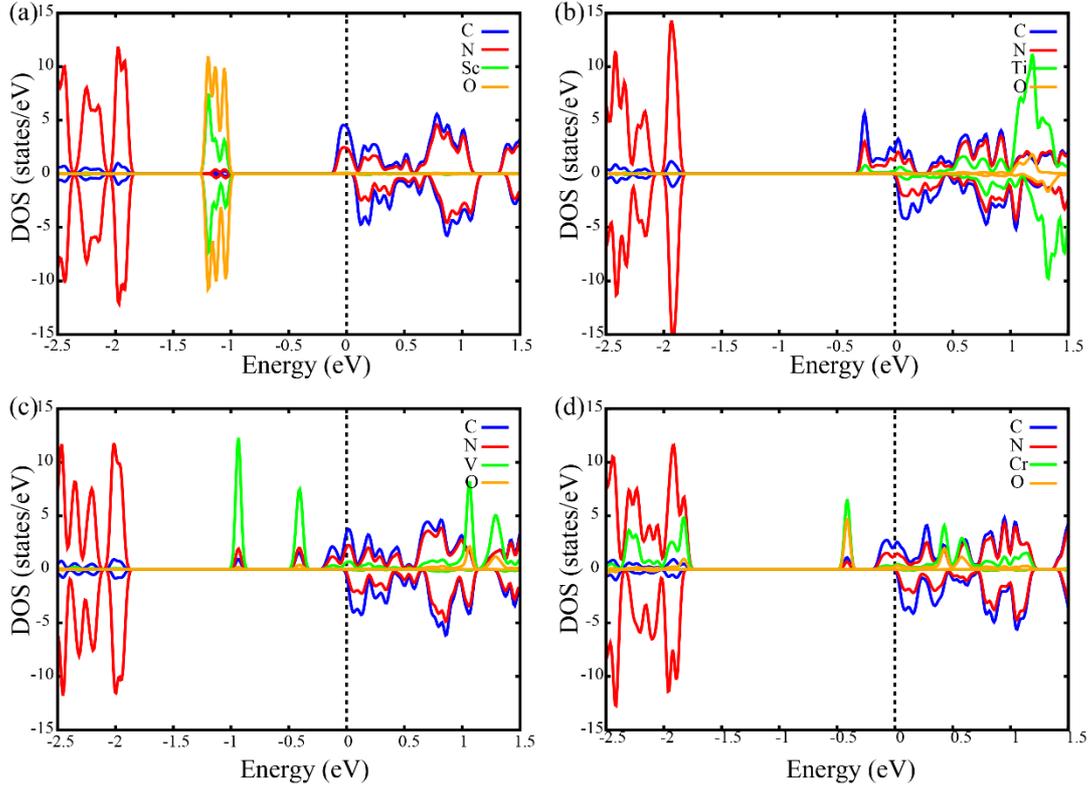

**Fig. 2** Density of states of the C₃N₄-TM-O system, where TM= (a) Sc, (b) Ti, (c) V, (d) Cr.

To explore the magnetic ground state, ferromagnetic and antiferromagnetic configurations were considered by using a $2\times2$ supercell of C₃N₄–TM–O systems. The energies of these differently coupled systems were computed to calculate the exchange energy ($E_{xc} = E_{FM} - E_{AFM}$). As shown in Table 2, the calculated exchange energies $E_{xc}$ are -27, -11, -32, -30, and 7 meV, respectively. The results indicate that Sc, Ti, V, Cr atoms are coupled ferromagnetically with each other, whereas Mn exhibits an antiferromagnetic coupling. The calculated magnetic moments of all the C₃N₄–TM–O systems are 1, 2, 3, 4, and 5 μB per unit cell for TM = Sc, Ti, V, Cr, and Mn, respectively.

Having studied the magnetic ground state of C₃N₄–TM–O, to investigate the electronic properties and the possible half-metallic property around the Fermi level of the molecular TM–O embedded 2D nanosheet, we also investigated the density of states (DOS) of all the ferromagnetic systems, as shown in Fig. 2. Interestingly, we found that all the ferromagnetic systems exhibit a half-metallic property. For the system of C₃N₄–Sc–O, the Sc atom has little contribution to magnetism for the symmetry of DOS. However, for C₃N₄–TM–O (TM = Ti, V, Cr) systems, the TM atoms have an obvious influence on the magnetism. Besides, we also investigated the charge transfer process by performing Bader charge analysis[43-44], as shown in Table 2. There are 0.65, 0.83, 0.75, 0.65, and 0.57 $e$ transferring from TM atoms to the g-C₃N₄ nanosheet for C₃N₄–TM–O (TM = Sc, Ti, V, Cr, Mn) systems, respectively. The transferred electrons occupy the conduction band of the g-C₃N₄ nanosheet and induce finite magnetic moments at Fermi energy.

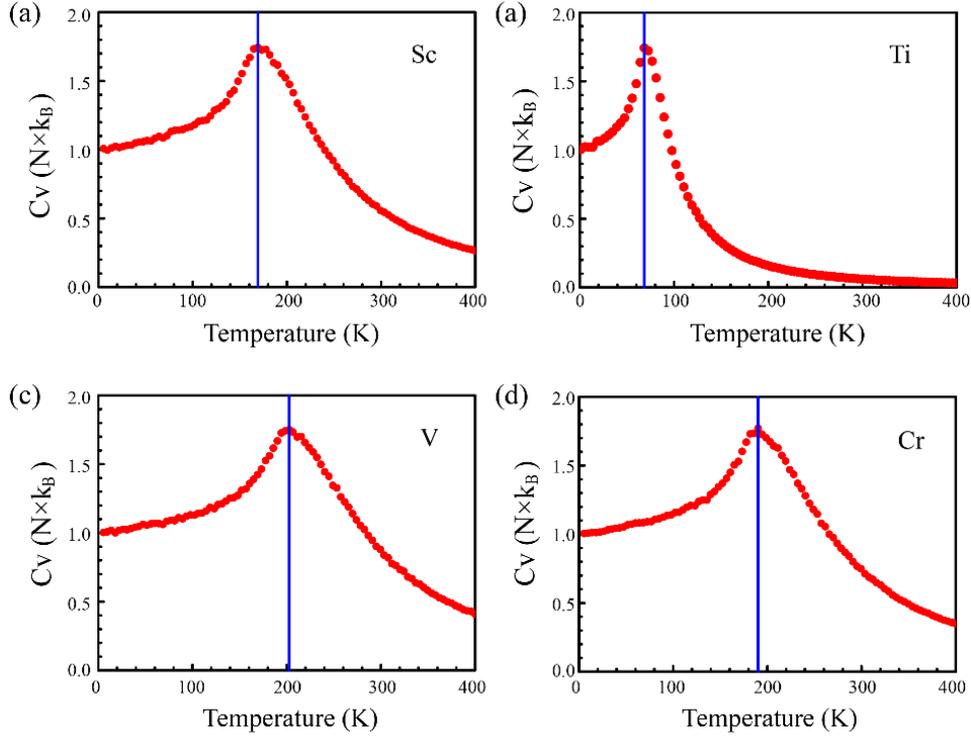

**Fig. 3** Heat capacity per spin (Cv/N•$k_B$) of $C_3N_4$-TM-O system which diverges at the Curie temperature, where TM = (a) Sc, (b) Ti, (c) V, (d) Cr.

The magnetic behavior under finite temperatures was also investigated so as to facilitate the use of ferromagnetic nanosheets in various applications. For the present study, Monte Carlo simulations using the Heisenberg model were performed to estimate the Curie temperature ($T_c$)[39-40, 45]. The heat capacity per spin (Cv/N•$k_B$) was obtained with respect to the temperature of all the systems, and it is plotted as in Fig. 3. The heat capacity peaks at the critical temperature where the phase transition occurs, indicating that the $T_c$ of the $C_3N_4$–TM–O (TM=Sc, Ti, V, Cr) framework are 169, 68, 203, and 190 K, respectively. The $T_c$ values are comparable with the high $T_c$ of some 2D half-metallic materials, which have recently been demonstrated based on first-principles calculations[46-48].

**Table 2**. Exchange energy ($E_{xc} = E_{FM} - E_{AFM}$), and Bader charge transfer from TM atoms to g-$C_3N_4$ nanosheet.

| Metal oxide | Sc-O | Ti-O | V-O | Cr-O | Mn-O |
|---|---|---|---|---|---|
| $E_{xc}$ (meV) | -27 | -11 | -32 | -30 | 7 |
| Bader charge (e) | 0.65 | 0.83 | 0.75 | 0.65 | 0.57 |

### 3.3 Half-metallic origin of $C_3N_4$–TM–O

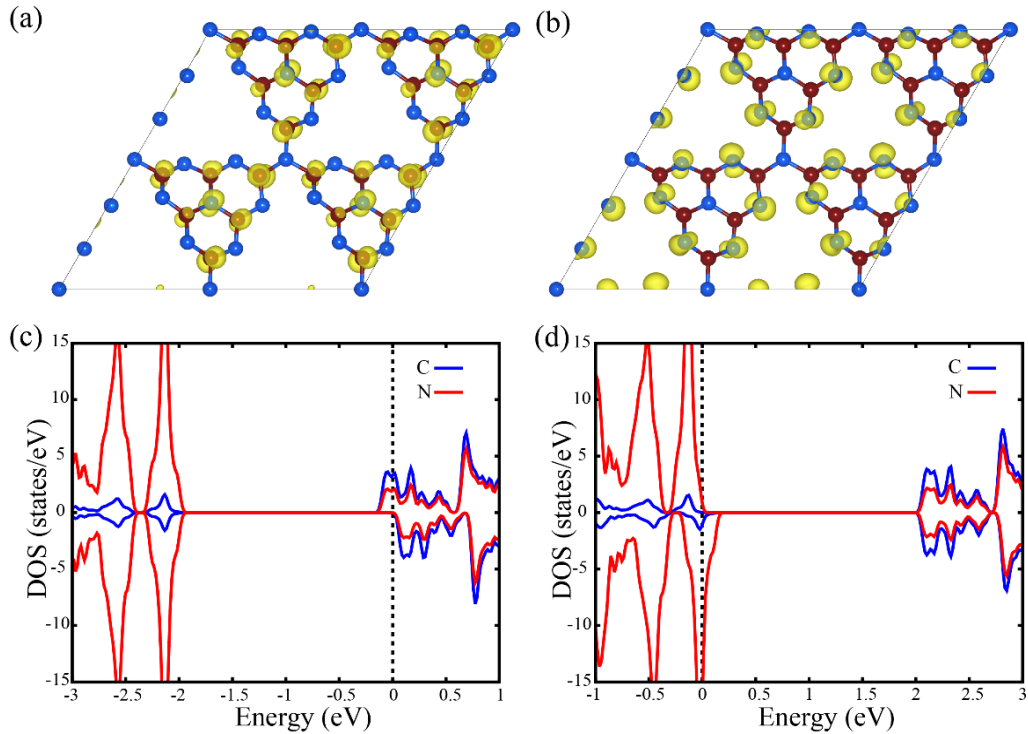

**Fig. 4** Spin polarized electron density difference (0.001 eV/Å³) of g-C$_3$N$_4$ with (a) an electron injected and (b) a hole injected. PBE calculated density of states (DOS) of g-C$_3$N$_4$ with (c) an electron and (d) a hole injected. The dotted vertical lines at 0 eV represent the Fermi energy.

Foreign atoms doped into the g-C$_3$N$_4$ system may inject the valence electrons or holes, which will certainly influence the electronic and magnetic properties of g-C$_3$N$_4$. As mentioned, all the C$_3$N$_4$–TM–O systems exhibit charge transfer between TM atoms and the g-C$_3$N$_4$ nanosheet. To clarify the origin of the half-metallicity, we examine the electronic structures of g-C$_3$N$_4$ by injecting one electron and hole, respectively. The spatial distribution of spin-polarized electron density of the two doping configurations calculated from the charge density difference between the spin-up and spin-down channels ($\Delta\rho = \rho_\uparrow - \rho_\downarrow$) is shown in Fig. 4(a) and (b). It is clear that both doping configurations have spin-polarized ground states. The $\Delta\rho$ isosurfaces of the two configurations display different features: The electron spin polarization occurs at the carbon atoms and the central nitrogen atom for the configuration with an injected electron, whereas it occurs at the edge nitrogen atoms for the configuration with an injected hole. Fig. 4(c) and (d) represent the PBE-calculated DOS of the g-C$_3$N$_4$ nanosheet with an electron and a hole injected, respectively. For the electron-doped system, the transferred electron occupies asymmetrically on the two opposite spin directions of the conduction band of the g-C$_3$N$_4$ sheet, inducing a half-metallic property. For the hole-doped system, the half-metallicity can be understood by the fact that the pair of electrons is broken up in different spin channels in the valence band. The C$_3$N$_4$–TM–O system is actually an example of injecting electrons into the g-C$_3$N$_4$ nanosheet. Injecting electrons and holes may be potential ways to introduce half-metallicity for 2D g-C$_3$N$_4$ nanosheets.

## 4 Conclusions

In summary, we systematically studied the electronic and magnetic properties of the molecular TM–O (TM = Sc, Ti, V, Cr, Mn) embedded in 2D g-$C_3N_4$ nanosheets. The negative binding energies indicate that the TM–O binds to the g-$C_3N_4$ nanosheet strongly. The results suggest that the 2D $C_3N_4$–TM–O framework is ferromagnetic for TM = Sc, Ti, V, Cr, but antiferromagnetic for TM = Mn. All the ferromagnetic systems exhibit a half-metallic property. Furthermore, Monte Carlo simulations based on the Heisenberg model suggest that the Curie temperatures ($T_c$) of the $C_3N_4$–TM–O (TM = Sc, Ti, V, Cr) framework are 169, 68, 203, and 190 K, respectively. Based on Bader charge analysis and electron injection, we found that both electrons and holes can induce half-metallicity for g-$C_3N_4$ nanosheets. The half-metallic origin of $C_3N_4$–TM–O systems can be partially attributed to the transfer of electrons from TM atoms to the g-$C_3N_4$ nanosheet. This study provides theoretical insight to understand the origin, and subsequent achievement of half-metallicity for graphitic carbon nitride.

## Acknowledgements

This work was partially supported by the National Natural Science Foundation of China (Grant Nos. 21773124 and 21503201), the Doctoral Fund of Ministry of Education of China (Grant No. 20120031120033), the Research Program for Advanced and Applied Technology of Tianjin (Grant No. 13JCYBJC36800), the Fok Ying Tung Education Foundation (No. 151008), the President foundation of the China Academy of Engineering Physics (Grant No. YZJJLX2016004), the National Key Research and Development Program of China (Grant No. 2016YFB0201203), and the Special Program for Applied Research on Super Computation of the NSFC-Guangdong Joint Fund (the second phase) under Grant No. U1501501. We appreciate the support from the National Super-Computing Center at Tianjin, Wuxi, and Guangzhou. S.L.H. is grateful to the foundation by the Recruitment Program of Global Youth Experts.